\documentclass[superscriptaddress,groupedaddress,nofootnoteinbib,12pt]{article}  % for review and submission

\usepackage{color}
\usepackage{graphicx}  % needed for figures
\usepackage{dcolumn}   % needed for some tables
\usepackage{bm}        % for math
\usepackage{jcappub}

\def\be{\begin{equation}}
\def\ee{\end{equation}}
\def\ba{\begin{eqnarray}}
\def\ea{\end{eqnarray}}
\def\beq{\begin{eqnarray}}
\def\eeq{\end{eqnarray}}

\def\mpl{M_{\rm Pl}}

\def\E{\mathcal{E}}

\def\d{\mathrm{d}}

\def\L*{{\cal L}_*}
\def\L{\mathcal{L}}
\def\({\left(}
\def\){\right)}

\def\nn{\nonumber}

\def\<{\langle}
\def\>{\rangle}

 \def\neq {\not\equiv}

\def\cs2{c_{s}^{2}}

 \def\ep{\varepsilon}

 \def\be   {\begin{equation}}   \def\ee   {\end{equation}}

 \def\ba  {\begin{eqnarray}}   \def\ea  {\end{eqnarray}}

\begin{document}

\title{Point particle motion in topologically nontrivial space-times}
\author{Andrew Matas,} 
\author[1]{Daniel M\"uller,\note{On leave of absence from Instituto de F\'\i sica Universidade de Bras\'\i lia, Caixa Postal 04455, 70919-970, Bras\'\i lia, Brazil}}
\author{and Glenn Starkman}

\affiliation{CERCA/Department of Physics, Case Western Reserve University, 10900 Euclid Ave, Cleveland, OH 44106, USA}

\abstract{It is well known that compactifying a space can break symmetries that are present in the covering space. In this paper we study the effects of such topological symmetry breaking on point-particle motion
when the particle is coupled to a massless field on the space.
 For a torus topology where Lorentz invariance is broken but translation invariance is maintained, particles can move at a constant velocity through the space;
however,  non-local, velocity-dependent forces arise whenever the particle is accelerated. 
For a  topology where translation invariance is broken, such as the Klein bottle,
 interactions  with the massless field generate an effective potential as a function of position. 
The potential creates special stable points in the space, 
and prevents  constant velocity motion.  This latter would appear to be the generic case.
This class of effects may be applicable whenever a localized object moves through a compactified bulk, 
such as in brane-world cosmology, or some condensed matter systems.}

\maketitle

\section{Introduction}

Topologically non-trivial spacetimes are common in modern theoretical physics.
They have featured in attempts to unify gravity with other forces since \cite{Nordstrom:1988fi}
(but are most often associated with Kaluza \cite{Kaluza:1921tu} and Klein \cite{Klein:1926tv,Klein:1926fj}).
This includes, of course, string theory and its generalizations,
where extra spatial dimensions are required 
for mathematical consistency of the theory \cite{Lovelace:1971fa} 
and must be hidden (or mostly hidden \cite{Dienes:1998vh}) from observation -- 
typically by compactification at a very short distance scale.  
It also includes large extra-dimensional solutions to the gauge hierarchy problem in which the extra
dimensions are hidden by compactification
\cite{Antoniadis:1990ew,ArkaniHamed:1998rs,Kaloper:2000jb}.
Additionally, one or more of the three large spatial dimensions in our universe can be compact, 
even if the curvature is flat (first proposed by \cite{Schwarzschild}; 
for more recent reviews see for example \cite{Luminet:2013ama,Starkman:1998qx}), 
although there are strict constraints on this possibility 
(most recently and strictly: \cite{Vaudrevange:2012da,Ade:2013vbw}).
\\

A common type of compact space is formed by taking an infinite manifold and modding out by a discrete subgroup of the full isometry group. For example, a cylinder $R\times S^1$ may be viewed as a taking the Euclidean plane $R^2$ and identifying $x$ with $x+L$, thereby modding out by a discrete subgroup of translations. Compactifiying a space in this manner can break symmetries that are present in the uncompactified space. A space-time with cylindrical spatial slices is not Lorentz invariant, because the slices do not remain purely spatial under a boost along the compact direction, even though the metric is locally Minkowski. The loss of symmetry can have physical implications.
\\

A famous example is the twin paradox with a compact extra dimension 
(see for example \cite{Peters:1983as,Barrow:2001rj,Barrow:2003ma,Uzan:2000wp}). 
Consider twins, one of  whom remains on Earth 
while the second  moves at a speed close to the speed of light around the compact direction. 
Direct computation shows, in the usual way of the twin paradox, 
that the moving twin returns younger than the stationary twin.
However, the moving twin Twill return to the first twin without ever having accelerated, 
so we cannot appeal to the standard resolution of the twin paradox to explain the apparent
violation of the relativity of motion.
The new resolution is that there is a topological difference between the twins: 
the worldlines of the two twins have different winding numbers. 
In other words, there is a preferred frame in which the compactified dimension is purely spatial,
and in which only the moving  twin winds around the space. 
\\

Compactificiation can have other implications.
In a generic compactification, 
many symmetries of the covering space may be broken beyond just Lorentz invariance. 
For example, the Klein bottle breaks translation invariance 
even though the space continues to admit a flat metric.  
The seminal work of de Witt et al in \cite{DeWitt:1979dd} 
established that the vacuum (Casimir) stress energy tensor is not Lorentz invariant,
and in the Klein bottle is not translation invariant.  
This perhaps inspire a considerable effort to calculate the vacuum stress energy 
in a variety of manifolds (eg. \cite{Muller:2001pe,Lima:2006rr}).  
This Casimir stress energy can have implications 
on the evolution of the space itself \cite{Nasri:2002rx} and
or on the motion of branes in the space \cite{Greene:2011fm,Jacobs:2012id,Jacobs:2012ph}. 
Loop corrections can also change the spectrum of the Kaluza-Klein tower, 
which was fixed at tree level by the 5 dimensional Lorentz invariance \cite{Cheng:2002iz}. 
As a result, it is important to understand these effects when dealing with compactified spaces.
\\

In this work we will be mainly focused on the dynamics of 
a point particle coupled to massless fields living in a topologically non-trivial space. 
There is a long tradition of studying radiative effects on the motion of point particles. 
The classic studies  by  Lorentz and Abraham \cite{Lorentz:1904,Abraham:1905} 
were followed up  by Dirac \cite{Dirac:1938nz} and then by many, but especially Rohrlich and collaborators
(see for example \cite{Rohrlich:1964,Rohrlich:2007}. 
This work was extended to curved space in 1960 by de Wiit and Brehme \cite{DeWitt:1960fc}, 
and more recently by many others \cite{Mino:1996nk,Quinn:1999kj,Galley:2005tj,Galley:2006gs};
for a review, see \cite{Poisson:2011nh}. 
Radiation reaction has also been studied 
in space-times with other than $3+1$ dimensions \cite{Galtsov:2001iv,Kazinski:2002mp} 
and in wormhole \cite{Khusnutdinov:2007wq,Krasnikov:2008kr,Bezerra:2010zz,Bezerra:2009wj} 
and cosmic string \cite{Linet:1986yp,Khusnutdinov:1995qr,Muniz:2013iua,Krtous:2006fb} space-times. 
However as far as we know, point particle motion has not been studied in detail in 
spacetimes with non-trivial spatial topology.
\\

The point particle can serve as a toy model for a brane, 
or for an ordinary quantum particle, moving through space.  
By focusing on point-particle motion as opposed to the Kaluza-Klein spectrum, 
we can describe ``localized" effects that are harder to describe in the Kaluza-Klein picture. 
As an example application, 
it is known that in brane-world cosmology scenarios the brane can radiate bulk gravitons 
\cite{Barvinsky:2002kh,Remazeilles:2008sg,Durrer:2007ww,Ruser:2007wz,Leeper:2003dd,Langlois:2002ke,Langlois:2003zb,Cartier:2005br} 
(for a review of brane-world gravity see \cite{Maartens:2003tw}). 
We imagine that on a topologically non-trivial space the radiated gravitons would interact with the 
emitting brane at later times on the brane's worldline.
\\

There are dynamical effects due to the topology 
that can change the motion of the point particle moving in a compact space 
relative to motion in the covering space. 
Because there are modes in the fields with which the particle interacts 
that have non-zero winding number (that is, field modes that wrap around the space), 
a point particle in a topologically non-trivial space can interact with itself. 
This will appear as an effective non-local interaction for the point particle. 
Since the particle can radiate into the bulk, the non-local interactions can 
result in dissipation. 
We will study the dynamics of these self-interactions for different topologies of flat space.
\\

Our main results are
 \begin{itemize}
  \item A point particle in a Klein bottle  sources a static potential with which it itself interacts. Consequently, the particle worldline will not be a geodesic of the spacetime, even in the absence of an external force. Furthermore, because translation invariance is broken, there are local minima in this potential -- stable locations in the manifold where the particle will settle since it can dissipate energy and momentum by emission of excitations of the massless fields with which it couples. 
 \item Even for the torus topology, which preserves translation invariance, 
 there are \emph{velocity-dependent} forces that react on the particle if it is accelerated. 
 The particle can emit radiation, and later interact with that radiation. 
 The effect is non-local from the perspective of the particle. 
 The particle can also dissipate energy that radiates to infinity.

\end{itemize}

The rest of this paper is organized as follows. 
In Section \ref{sec:review} we review field theory on a compact manifold and set up our formalism. 
In Sections \ref{sec:klein-bottle} and \ref{sec:torus}, 
we study point-particle motion on a Klein bottle and on a torus respectively.  
In both cases we focus for simplicity on manifolds with even space-time dimension. 
We work in units where $\hbar=c=1$ and
use a ``mostly plus'' metric convention, $\eta_{\mu\nu} = {\rm diag}(-1, 1\cdots 1)$.

\section{Review of field theory on topologically non-trivial manifolds}
\label{sec:review}
We first review some basic facts about field theory in compact spaces.

\subsection{Compactifications of Minkowski space}

In this work we consider compactifications of d-dimensional Euclidean space $E^d$
(in two dimensions, the Euclidean plane, $E^2$)
obtained by modding  by a discrete subgroup $\Gamma_A$ (where $A$ runs over all groups)
of the  isometry group of $E^d$.  
These  are the crystallographic groups in $d$ dimensions.
Meanwhile, we consider time to be uncompactified.
Thus we are interested in space-time manifolds of the form $R\times E^d/\Gamma_A$.

Our group elements $\gamma_n\in\Gamma_A$ (for some particular $A$) thus identify
\be
\label{eqn:isometry}
(t, x^i) \sim (t, \gamma_n x^i).
\ee
The identification (\ref{eqn:isometry}) breaks Lorentz invariance 
by singling out a preferred frame in which $\gamma_n$ acts purely on the spatial coordinates. 
In other words, 
all inertial frames where the identification conditions are simultaneous are related by
translations and rotations, not boosts.

In keeping with tradition,  we shall only consider compactifications that give rise to manifolds,
although it is certainly possible to consider more general compactifications.
This limits $\Gamma_A$ to ``freely acting'' subgroups -- those with no fixed points under 
subroup elements (other than the identity).  In other words
$\gamma_n x^i \neq x_i$ for any $x_i$ in the space and for any $\gamma_n \in\Gamma_A$.
In 1910-1912, Bieberbach showed \cite{Bieberbach:1911,Bieberbach:1912} 
that there are only a small number of freely acting crystallographic groups 
for a Euclidean space of given dimension. 
For example, for $3$-dimensional Euclidean space there are only ten compact manifolds.

In this work we focus on field configurations that are strictly periodic\footnote{
	It is also possible to consider twisted field configurations, 
	where $\phi_a(x) = M[\gamma]_{ab}\phi_b(\gamma x)$. 
	See for example \cite{DeWitt:1979dd}. We will not consider this possibility here.}. 
In other words, if we identify $x \sim \gamma x$, then
\be
\phi_a(x) = \phi_a(\gamma x),
\ee
where $a$ is a generic index that includes internal and space-time indices. This will allow us to write any field solving a linear equation of motion in terms of the covering space as 
\be
\label{eq:field-expansion}
\phi_{\mathcal M}(x)=\sum_i \phi(\gamma_n x),
\ee
as long as this summation converges. The $\gamma_n$ are the elements of the group which defines the geometric space. This is called the method of images and is used very often in harmonic analysis \cite{Camporesi:1990wm}. For a recent book see \cite{Terras:2013}. 

For compactifications of Euclidean-$n$ space the summation \ref{eq:field-expansion} can be regrouped into separate summations over the torus. This fact was used previously to obtain the Casimir vacuum energy for the flat oriented closed spaces \cite{Lima:2006rr}. 

\subsection{Massless Propagators}

 We are interested in purely classical physics that is described by the retarded propagator. 
 For a massless scalar field $\phi$, the retarded propagator is a solution to
\be
\square_x G_{\rm ret}(x,x') = \delta^{(D)} (x - x'),
\ee
with the boundary condition that $G(x,x') = 0$ for $x^0 < x'^0$.

On Minkowski space-time, rotational and translational invariance 
allow us to express the Green's function as a function of 
$T \equiv t - t'$ and $R \equiv |\vec{x} - \vec{x}'|$ for $D>2$ as
\ba
\tilde{G}_{\rm ret}^{(D)}(T,R) &=& \frac{1}{4\pi} \left[ \frac{1}{2\pi R} \partial_R \right]^{(D-4)/2} \frac{\delta(T-R)}{R}, \ \ D\ {\rm even}, \nn \\
\tilde{G}_{\rm ret}^{(D)}(T,R) &=& \frac{1}{4 \pi} \theta(T) \left[ \frac{1}{2\pi R} \partial_R \right]^{(D-3)/2} \frac{\theta(T-R)}{\sqrt{T^2 - R^2}}, \ \ D\ {\rm odd}.
\ea

The presence of a tail in the Green's function 
except in flat unperturbed space-times of even dimension is well known. 
Intuitively, the tail can be understood by considering (see for example \cite{Galtsov:2001iv})
the delta function source in odd space-time dimensions 
as being a line of point sources in even dimensions. 
The radiation from different points on the line will arrive at different times, leading to the tail. 
See also \cite{Chu:2011ip} and \cite{Dai:2013cwa} for an alternative perspective on the tail. 
For simplicity alone, we will focus on unperturbed flat even space-time dimensions.

In a general topology, 
we can express the retarded propagator $G_{\rm ret}$ 
in terms of the Minkowski space propagator $\tilde{G}_{\rm ret}$ 
by using Equation \ref{eq:field-expansion}:
\be
\label{eq:general-greens-function}
G^{(D)}_{\rm ret}(x,x') = \sum_n \tilde{G}^{(D)}_{\rm ret}(x, \gamma_n x').
\ee
We will use this as a starting point when we consider specific topologies below.

\subsection{Effective Action for a Point Particle}
\label{sec:effective-action}

We will consider a generic  action describing 
the motion of a point particle coupled to some bosonic field:
\be
S[X,\phi] = S_p[X] + S_\phi [\phi] + \int d^4 x \phi J,
\ee
where
\ba
S_p[X] &=& m \int \d\tau, \nn \\
S_\phi[\phi] &=& \int \d^4 x -\frac{1}{2} (\partial \phi)^2, \nn \\
J(x,X) &=& Q \int \d \tau \delta^{(4)} \left(x - X(\tau)\right).
\ea
The breaking of various symmetries of the covering space 
is encoded in the boundary conditions of the integral. 
For example, in the torus case, 
where invariance under boosts in the compact direction $z$ is broken, 
there is only one frame where the limits of the $z$ integral are 0 to $L$.

We may write the effective action for $X$ after integrating out $\phi$ as
\be
e^{i S_{\rm eff}[X]} = \int \mathcal{D} \phi e^{i S_p[X] + i S_\phi[\phi] + i  \int d^4 x \phi J}.
\ee
The integral over $\phi$ is Gaussian and can be performed easily, yielding
\be
\label{eq:effective-action}
S_{\rm eff}[X] = S_p[X] + \frac{Q^2}{2} \int \d \tau \int \d \tau' \sum_n \tilde{G}_F \left( X(\tau), \gamma_n X(\tau') \right),
\ee
where $\tilde{G}_F$ is the Feynman propagator. This can be easily generalized to fields with arbitrary spin and to multiple fields.

We may interpret the sum over $n$ as giving the contribution of different winding sectors of the theory:
\begin{itemize}
\item The $n=0$ term is divergent. This corresponds to one loop corrections that would be present in Minkowski space. For example, these divergences contribute to the renormalization of the point particle's mass and, in $D=4$, results in the famous Abraham-Lorentz-Dirac equation of motion for the classical point particle. 
\item The $n\ne0$ terms represent the effects of the nontrivial topology. These are extra finite contributions to the effective action for $X$. The contributions are non-local because we have integrated out $\phi$. The symmetries are now manifestly broken because of these terms.
\end{itemize}
The advantage of writing things in this way is that we can express the motion of the point particle in terms of the basic scattering amplitude $G(x_1, x_2)$.

We work in real space, instead of Fourier space,
focusing on  winding modes rather than Kaluza-Klein modes. 
This is because we want to study the motion of a localized point particle in the manifold.
Below we compute corrections to the point particle equations of motion 
in Klein bottle and torus topologies. 
These are ultimately due to the above corrections to the effective action, 
which are a measure of scattering off of winding modes. 

\section{Static potential for the Klein bottle}
\label{sec:klein-bottle}

We now consider a flat space-time $\d s^{2}=-\d t^{2}+\d\vec{x}^{2}$, 
with $D\geq 4$ space-time dimensions, where the
spatial slices have the topology of a Klein bottle times $R^1$. The importance of the Klein bottle is that any
two-dimensional section of a compact Euclidean $3-$manifold is either a
$2-$torus or a Klein bottle.

Unlike the $2$-torus,  the Klein bottle topology breaks translation invariance.
This is the generic situation among compactifications. 
It means that a single charged particle in this space will give rise to an
inhomogeneous static potential, 
with which the particle can itself interact. 
The trajectory of a moving particle consequently will not  be a geodesic of the metric,
and, since the particle can dissipate energy  by the emission of  massless scalars, 
it will settle into a local minimum of this potential.
This is a purely classical effect, 
making it distinct from the effects studied in \cite{Jacobs:2012id,Jacobs:2012ph}. 
The charged particle coupled to a massless scalar field
can also be considered a qualitative model for a brane coupled to a bulk field, 
such as a Kalb-Ramond field or the gravitational field.

\begin{figure}[htpb] 
\begin{center}
\includegraphics[scale=0.6]{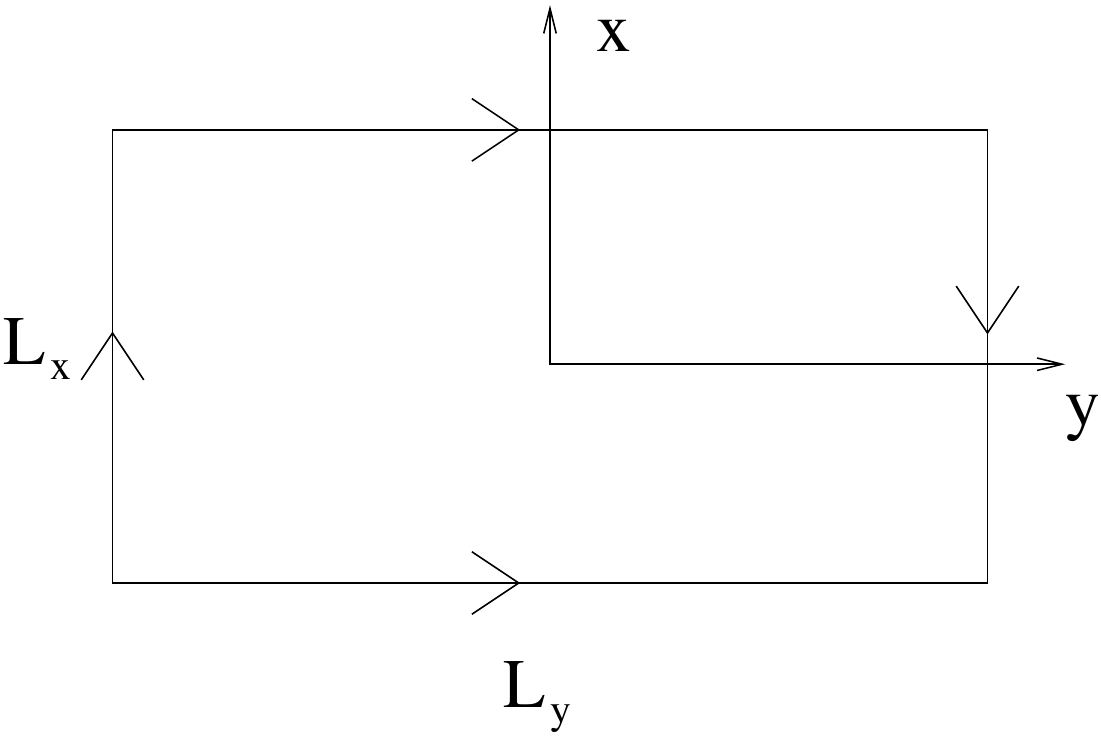} \caption{Fundamental region of Klein bottle.
\label{K}}
\end{center}
\end{figure}

\begin{figure}[htpb] 
\begin{center}
\includegraphics[scale=0.6]{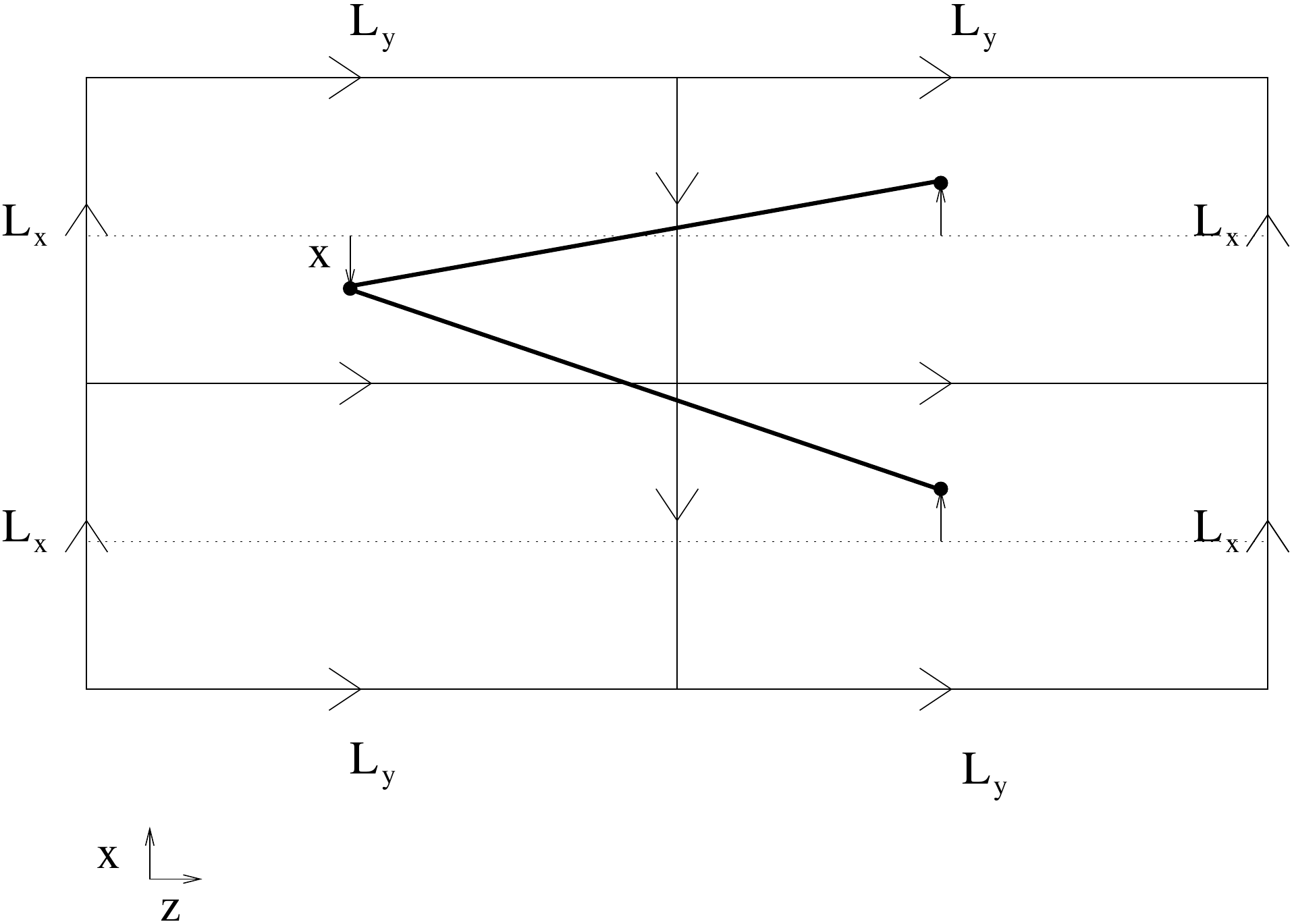} 
\caption{Fundamental region of Klein bottle together with other $3$ copies
of it. When $x=0$ we have the same potential as when $x=\pm L_x/2$.
Increasing the distant by gradually moving the charge in the $x<0$
direction it can be seen that there will be a minimum between $x=0$
and $x=-L_x/2,$ when the charge is going to be equidistant from the
to other images shown. 
\label{charge} }
\end{center}
\end{figure}

The Klein bottle topology is defined by the following identifications of the Euclidean plane:
\be
\left(
\begin{array}{c}
t \\
x \\
y
\end{array}
\right)
\sim
\gamma_{n,m}
\left(
\begin{array}{c}
t \\
x \\
y
\end{array}
\right)
=
\left(
\begin{array}{c}
t \\
 (-1)^m x+ nL_x \\
y + m L_y
\end{array}
\right)
\ee
The fundamental region for the Klein bottle is shown in Fig. \ref{K}. The top and bottom sides are identified as in the torus. The left and right sides are ``anti-identified", meaning that as we pass the lines $y=\pm\frac{L_y}{2}$ we send $x\rightarrow -x$. The Klein bottle is covered two times by the torus. 
In Fig. \ref{charge} we show a piece of the covering space. 
This figure helps to visualize the breaking of translation invariance
which generates the static potential. 

It can be seen visually that the potential on the line $x=0$ 
will be the same as the potential on the lines $x=\pm L_x/2$, 
and both correspond to a cubic type lattice,
for which the force is zero. 
Gradually moving the charge from position $x=0$ to $x<0$, 
as shown in Fig. \ref{charge}, 
it can be seen that there will be a minimum between $x=0$ and $x=-L_x/2,$
when the charge is going to be equidistant from the two other images
shown. We shall demonstrate this in the following.

\begin{figure}[htpb]
\includegraphics[scale=0.5]{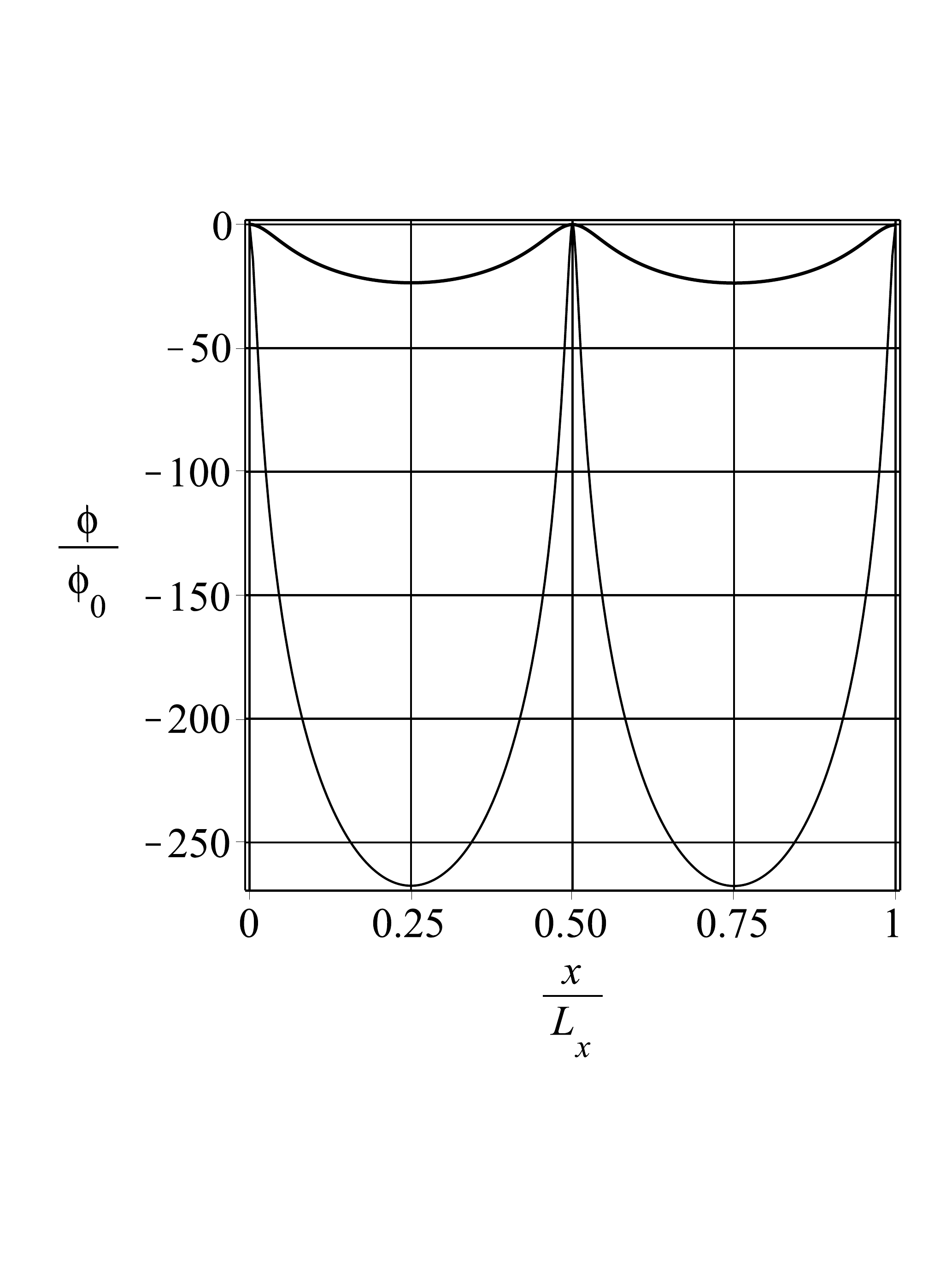} 
\caption{Retarded potential for a static charge in Klein's manifold in $D=4$ 
for two values of the dimensionless parameter 
$\beta \equiv \frac{L_y}{L_x}$: $\beta=0.05$ and $\beta=0.01$. 
An infinite constant is subtracted which corresponds to the potential at $x=0$.
The force is zero at $x/L_x=0$, $x/L_x=1/4$, $x/L_x=1/2$ and $x/L_x=3/4$, 
while $x/L_x=1$ is identified with $x/L_x=0$. 
The potential is smooth everywhere. 
It becomes deeper as $\beta$ decreases.
\label{potencial} 
}
\end{figure}

The holonomy group for the Klein bottle is a rotation by $\pi$ around the $y$-axis.
Its dimension is $2$, 
allowing the Green's function to be written as a sum of two pieces:
\ba
G(x,x')&=& \sum_{m,n} \tilde{G}\left(x, y,t; (-1)^m x + n L_x, y + m L_y ,t'\right)  \nn \\
&=& \sum_{m\ {\rm even},n} \tilde{G}\left(x,y,t; x + nL_x, y+mL_y,t'\right) + \sum_{m\ {\rm odd},n} \tilde{G}\left(x,y,t; -x + nL_x, y+mL_y,t'\right) \nn \\
&\equiv& G_1(x-x', y-y',t-t') + G_2(x+x',y-y',t-t').
\ea
For static sources, the Green's function will give rise to a potential. 
$G_1$ is translation invariant and yields an uninteresting (infinite) constant,
 which can be subtracted from the potential. 
 $G_2$ on the other hand breaks translation invariance in the $x$ direction 
 and so generates a non-trivial potential. 
 For a point charge with charge $q$ at the spatial position $(x,y)$, 
 the potential is given by
\ba
\phi(x) &=& \tilde{\phi} + q \sum_n \int\ \d\Delta t\ G_2\left(x+nL_x,mL_y,\Delta t\right) \nn \\
&=& \tilde{\phi} + \frac{q}{L_x^{D-3}}\sum_{\substack{m,n=-\infty\\(m,n)\neq0}}^{+\infty}\frac{1}{\left[\left(\frac{2x}{L_x}-m\right)^{2}+4n^{2}\left(\frac{L_y}{L_x}\right)^{2}\right]^{(D-3)/2}},
\label{potential}
\ea
where the direct path ($m=n=0$) is  excluded from the sum, 
and where $\tilde{\phi}$ is an arbitrary constant that we choose so that $\phi(0)=0$. 
Note that $\phi(x+\frac{NL_x}{2})=\phi(x)$ for integer $N$.

The derivative of the potential is
\[
\frac{\partial \phi}{\partial x}=-2(D-3)\frac{q}{L_x^{D-2}}\sum_{\substack{m,n=-\infty\\(m,n)\neq0}} \frac{\(\frac{2x}{L_x}-m\)}{\left[\left(\frac{2x}{L_x}-m\right)^2+4n^2\left(\frac{L_y}{L_x}\right)^2\right]^{(D-1)/2}}.
\]
The derivative can only be zero if there are cancellations term by term. This will occur precisely when $\frac{2x}{L_x}$ is an integer or half integer. When $\frac{2x}{L_x}$ is an integer the potential is at a maximum, when it is a half integer the potential is minimized.

In Fig. \ref{potencial} we plot $\phi(x)$ for $D=4$ 
and two values of the dimensionless parameter $\beta\equiv \frac{L_y}{L_x}$. 
The inhomogeneity in $\phi(x)$ increases as $\beta$ decreases.

\section{Point particle motion on the Torus}
\label{sec:torus}
We next consider point-particle motion on the $2-$torus. 
Because the $2-$torus does not break translation invariance,
it admits constant velocity solutions. 
However, there can still be effects from the breaking of Lorentz invariance.
To study these effects we imagine that we have a point particle moving through the compact direction. 
The point particle radiates because it is accelerated (by an external force). 
The radiation will travel around the compact direction, 
where it can then scatter off the point particle.  
We compute the force exerted by the radiation on the point particle,
and the resulting backreaction.

We choose to compute the force because 
it is a natural, gauge-invariant observable defined in the topology frame, 
and it allows us to consider stability. 
Of course, based on the arguments of section \ref{sec:effective-action}, 
underlying this calculation is a scattering amplitude;
the force can also be viewed as a measure of 
the strength of the scattering of the winding bulk particles.

The torus is defined by $\gamma_n \vec{x} = \vec{x} + nL \hat{z}$, 
where $\hat{z}$ is a unit vector pointing in the (compactified) $z$ direction. 
In other words,
\be
(t,x^i) \sim (t,x^i + nL \hat{z}^i).
\ee
We work in coordinates where $\hat{z}^i=\delta^{i,3}$.

This definition breaks manifest Lorentz invariance. 
To work in an arbitrary frame, 
we would need to promote $\hat{z}$ to a four-vector. 
Instead, we compute everything in the frame where $\hat{z}^0=0$. 
We call this frame the \emph{topology frame}.

It is useful to note that in this frame the effective length of the space is
\be
L^\prime\equiv\sqrt{1+\gamma^2 v_z^2} L.
\ee

\subsection{Radiation into a bulk scalar field}
Consider a point particle coupled to a scalar field in four space-time dimensions. 
We consider generalizing to fields of other spins 
and to other numbers of dimensions in section \ref{sec:generalizations}. 
The action is
\be
S = \int \d^4 x \left(-\frac{1}{2} (\partial \phi)^2 \right) + m \int \d \tau + Q \int \d^4 x \phi(x) J(x,X),
\ee
where
\be
J(x,X) = \int \d \tau \delta^{(4)}(x - X(\tau)).
\ee
The equations of motion are\footnote{
	When deriving the equations of motion, 
	it is important to remember to write the worldline integrals 
	in a way that is manifestly invariant under reparameterizations of $\tau$. 
	Thus, one should replace $\d\tau$ by $\d \tau \sqrt{-\dot{X}^2}$ before varying.}
\ba
\label{eq:scalar-eoms}
\square \phi &=& - Q J(x,X) \nn \\
m_{\rm eff}(\phi)\ddot{X}^\mu &=& -Q\left(\eta^{\mu\nu} + \dot{X}^\mu \dot{X}^\nu \right) \partial^\nu \phi\big|_{x = X}
\ea
where $m_{\rm eff}(\phi) = m + Q \phi$.
The scalar field equation can be formally solved
\be
\label{eq:phi-solution}
\phi(x) = \int \d\tau \ G(x, X(\tau))  \nn %\\
= - \frac{Q}{4 \pi}\sum_n \frac{1}{\sigma_\mu(\tau_{n})\dot{\sigma}^\mu(\tau_n)}\,,
\ee
where $\tau_{n}$ is the solution to
\be
\label{eq:retarded-time}
\sigma_{n,\mu}(\tau_n) \sigma^\mu_n(\tau_n)= 0
\ee
and 
\be
\label{eq:proper-time-eqn}
\sigma_n^\mu(\tau) \equiv x^\mu - X^\mu(\tau) - n L \hat{z}^\mu.
\ee
Substituting this expression for $\phi$ into the point-particle equation of motion 
gives the same result we would have obtained by varying \ref{eq:effective-action} directly.

Equation \ref{eq:retarded-time} is impossible to solve in general 
as it is a complicated implicit equation that depends on $X^\mu$. 
However, for the $2$-torus inertial motion is an exact solution, as is now shown.

\subsubsection{Existence of constant velocity solution}
The special property of the $2$-torus topology 
is that the compactification does not break translation invariance. 
The existence of solutions where the particle moves at a constant velocity in an arbitrary direction 
is guranteed by momentum conservation.

Explicitly, we seek a solution of the form
\be
X^\mu(\tau) = X_0^\mu + u^\mu \tau.
\ee
Then the condition \ref{eq:proper-time-eqn} becomes a simple quadratic equation for $\tau_n$
\be
\label{eq:proper-time}
\left[ u_\mu \tau_n  + \left(x_\mu-x_{0,\mu}\right) +  n L \hat{z}_\mu \right]^2 = 0,
\ee
which may be  solved trivially. 
Then $\phi$ is given by \ref{eq:phi-solution},
with $\tau_n$ given by the solution to the quadratic equation.

As a consistency check, one can verify that 
the right hand side of the point-particle equation of motion (\ref{eq:scalar-eoms}) vanishes. 
This involves differentiating the numerator, 
and evaluating everything on the path of the particle at the site of the ``zeroth image," 
$x^\mu = u^\mu \tau_0$. 
The result for the $2$-torus is
\be
\left(\eta^{\mu\nu} + \dot{X}^\mu \dot{X}^\nu \right)\partial_\nu \phi  = - \frac{Q^2}{4\pi} \sum_n \frac{nL}{\left( |n| L \sqrt{1+\gamma^2 v_z^2}\right)^3} (\hat{z}^\mu + (u\cdot \hat{z})u^\mu )=0.
\ee
Constant velocity is therefore a solution of the equations of motion. 
For the Klein bottle, 
the sum would not vanish, and constant velocity would not be a solution.

\subsubsection{Approximations}

Having established that constant velocity is a solution to the equations of motion, 
we examine the consequences of perturbations around that solution.

Consider the emission of radiation from the particle, 
with $k_{max}$ being the highest energy mode that is radiated. 
For example, 
this radiation might occur as a result of the sudden acceleration of the particle 
with characteristic time scale $k_{max}^{-1}$.

We assume that 
\be
k_{max} L'  \gg 1.
\ee

That is, so that we are in the radiation zone
we take the wavelength of the radiation to be small 
compared to the size of the space in the particle's rest frame, 

We also assume that 
\be
\frac{k_{max}}{m} \ll 1.
\ee
so the point particle that can be treated classically. 

Combining these, we learn that
\be
m L' \gg 1.
\ee

\subsubsection{Force due to winding modes}
We are now able to compute the force perturbatively. 
Consider perturbations to the constant velocity solution
\be
X^\mu = \bar{X}^\mu + \ep^\mu,
\ee
where
\be
\bar{X}^\mu = X_0^\mu + u^\mu \tau.
\ee
Proper time parameterization is chosen, $X^2 = -1$. 

To first order in $\epsilon$, the equation of motion for the point particle is
\ba
\label{eq:perturbed-particle-eom-scalar-case}
m_{\rm eff}(\phi(\bar{X})) \ddot{\epsilon}^\mu &=& (\eta^{\mu\nu} + \dot{\bar{X}}^\mu \dot{\bar{X}}^\nu) \partial_\nu \delta \phi 
+ \left(\dot{\bar{X}}^\mu \dot{\epsilon}^\nu + \dot{\epsilon}^\mu \dot{\bar{X}}^\nu\right) \partial_\nu \phi(\bar{X}) \,,
\ea
where
\be
\delta \phi = \phi(\bar{X} + \epsilon) - \phi(\bar{X}).
\ee

This equation is non-local in time (since we have integrated out the scalar field);
it is of the form
\be
\ddot{\epsilon}(\tau_0) = \sum_n F(\epsilon(\tau_n)),
\ee
where $\tau_n < \tau_0$ for all $n$ in the sum. 
Thus, we may evolve from $\tau_0$ to $\tau_0 + \delta \tau$ for small $\delta \tau$ 
provided we know $\epsilon(\tau)$ for all $\tau<\tau_0$. 
Given the physical picture we have in mind, 
we take $\epsilon(\tau) = 0$ for $\tau<0$, 
and then consider $\epsilon$ to be some specified function for 
$0 < \tau < \Delta \tau \sim 1/k_{\rm max}$.

The field will then take the form
\be
\delta \phi \sim \frac{A \ddot \epsilon}{L} + \frac{B \dot \epsilon}{L^2} + \frac{C \epsilon}{L^3},
\ee
for some $A,B,C$ which can depend on the background quantities. 
The expansion must start at $O(1/L)$ since the field should die off at infinity. 
The dependence on $\epsilon$ and its derivatives is fixed by dimensional analysis. 

Focusing on the leading term in $1/L$ lets us neglect the second term on the right hand side of Equation \ref{eq:perturbed-particle-eom-scalar-case}. Thus the equation of motion becomes
\ba
\label{eq:perturbed-particle-eom-scalar-case}
m_{\rm eff}(\phi(\bar{X})) \ddot{\epsilon}^\mu &=& (\eta^{\mu\nu} + \dot{\bar{X}}^\mu \dot{\bar{X}}^\nu) \partial_\nu \delta \phi 
\ea

The leading order piece in $1/L$ dominates at large distances. 
Working in a regime where $k_{\rm max} L \gg 1$, 
we can neglect the terms proportional to $B$ and $C$.
To this same order of approximation, 
we may ignore perturbations to the retarded time $\tau_n$, taking $\tau_n = \bar{\tau}_n$. 
Then
\be
\partial_\mu \delta \phi = -\frac{Q}{4\pi} \sum_n \frac{(\ddot{\epsilon} \cdot u) \partial_\mu \tau}{(\bar{X} \cdot u)^2} + \mathcal{O}\left(\frac{1}{L^2}, \epsilon^2\right).
\ee

We can now compute the corrected equation for the point particle motion,
expressing the physical 3-force in the rest frame of the topology. 
The 3-acceleration is related to the 4-acceleration by
\ba
a^i & \equiv & \frac{d^2 x^i}{d t^2} 
= \left(\delta_{ij} - \frac{\dot X_i \dot X_j}{(\dot X^0)^2}\right)\frac{\ddot X^j}{(\dot X^0)^2} 
=\frac{1}{\gamma^2}(\delta_{ij} - v_i v_j) \ddot \ep_j + \mathcal{O}(\ep^2).
\ea
We find the 3-force
\ba
\label{eq:scalar-force}
F^i &=& m_{\rm eff}(\phi) a_i(\tau_0) \nn \\
&=& 
- \frac{Q^2}{4\pi} \sum_n \frac{1}{|n|L\left(1+\gamma^2 v_z^2 \right)^{3/2}} 
\left[ (\vec{a}_n\cdot \hat{z}) + \gamma^2 (\vec{a}_n\cdot \vec{v})(\vec{v}\cdot \hat{z}) \right] \hat{z}^i,
\ea
where $a_n \equiv a(\tau_n)$, and $\tau_n$ is the retarded proper time, 
given by the solution of \ref{eq:proper-time}
\be
\tau_n = \tau_0 - |n| L \left( \sqrt{1+\gamma^2 v_z^2} - \frac{n}{|n|} \gamma v_z \right).
\ee

\subsubsection{Features of the force}

There are several features of (\ref{eq:scalar-force}) that merit comment. 
First, we note that the force depends on the velocity of the particle relative to the topology frame, reflecting the global breaking of Lorentz invariance. 
\\

Next there is the issue of stability. 
For the scalar case, it is easy to see that 
the force will always be in the opposite direction as the original acceleration. 
Thus the radiation tends to return the particle  to its original velocity, 
stabilizing the velocity particle.
\\

The magnitude of the force is independent of $n$. 
This may be surprising from the point of view of an observer in the topology frame. 
From the perspective of such an observer, 
the left-moving radiation has moved a different distance than the right-moving radiation. 
Thus an observer in that frame might have expected the radiation moving `against' the velocity 
to have a larger effect than radiation moving `with' the velocity.

However, while the topology frame is useful for understanding the periodicity conditions, 
it is misleading to estimate the size of the force in that frame. 
Instead imagine boosting into the rest frame of the particle. 
In this frame the size of the space is $\gamma L$, and the periodicity condition is
\be
(t,z) \sim (t + n \gamma v L, z + n \gamma L).
\ee
In this frame the magnitude of the force due to $\pm n$ is manifestly the same. 
We also see that the arrival times of the radiation will be different. 
This effect was noted, for example, in \cite{Greene:2011fm}.
\\

Perhaps more surprisingly, 
the force is independent of the \emph{sign} of $n$. 
This is a consequence of the fact that the acceleration picks out a preferred direction. 
More explicitly, the field is given by
\be
\phi(x) = \frac{q}{4\pi}\left[\frac{1}{(x-\bar{X})\cdot \bar{X}} - \frac{(x-\bar{X})\cdot \dot{\epsilon}}{(x-\bar{X})\cdot \bar{X}} + \frac{\dot{\bar{X}}\cdot \epsilon}{((x-\bar{X})\cdot \bar{X})^2} + O(\epsilon^2)\right]
\ee
The dominant term is the second one, which is the radiative part of the field (since it is the piece for which $\partial \phi \sim \ddot{\epsilon}$).

In the rest frame of the charge, this piece has the form
\be
\phi_{\rm rad,rest\ frame}(x) 
= -\frac{q}{4\pi}\frac{(\vec{x}-\vec{\bar{X}}) \cdot {\dot{\vec{\epsilon}}}}{|\vec{r}|}.
\ee
This is clearly not spherically symmetric. 
In particular, 
the dipole form of the field 
guarantees that the force will be in the same direction 
for observers at $\vec{r}$ and $-\vec{r}$.

As a proof of principle, 
in section \ref{sec:time-dependent-coupling} 
we consider a case where the radiative part of the field is spherically symmetric. 
For a point particle this is necessarily artificial, 
but it could be considered to be a toy model for more realistic setups.
\\

Another interesting feature is that the force is entirely in the $\hat{z}$ direction. 
As we will show below, this is a special feature of the scalar field. 
\\

Finally, we comment on the size of the effect. 
The force creates an acceleration of the particle of the schematic form (ignoring relativistic effects)
\be
a(\tau) \sim \frac{Q^2}{ML}\sum_n \frac{a(\tau_n)}{|n|},
\ee
where $Q$ is a parameter measuring the strength of the interaction.
Thus the particle experiences a series of kicks as the radiation hits the particle. 
An individual kick has a size approximately given by 
\be
\delta v_{\rm single\ kick}\ \sim \frac{Q^2}{M L}.
\ee
Within the regime of validity of our calculation, 
this will be small (assuming $Q$ is small enough that we can trust perturbation theory).

The sum formally diverges in the limit $|n|\rightarrow \infty$. 
This corresponds to waiting a long time. 
Thus, though an individual pulse may give a small effect, 
over a long time the accumulated effect of the pulses can accumulate to a large effect on the trajectory.  
This is a special feature of $3+1$ dimensions.

Since the effect is linear in the number of light fields in the bulk, the effect of one kick can also become large if a large number of fields are present.

\subsubsection{Special cases}

Equation \ref{eq:scalar-force} is rather complicated, 
so we will consider two special cases to gain insight. 

\paragraph{Special Case 1: $\vec{v}\ ||\ \vec{z}$} \ \\

In this case we write $\vec{v} = v \hat{z}$. 
Then $\sqrt{1+\gamma^2 v_z^2}=\gamma$. 
This leads to
\be
F^i = - \frac{Q^2}{4\pi} \sum_n \frac{\vec{a}_n\cdot \hat{z}}{|n| \gamma L}\hat{z}^i.
\ee
which is suppressed by $|n|\gamma L$, the distance between the charges in their rest frame.

\paragraph{Special Case 2: $ \vec{v}\ \bot\ \hat{z} $} \ \\
Now  $\sqrt{1+\gamma^2 v_z^2}=1$. As a result,
\be
F^i = - \frac{Q^2}{4\pi} \sum_n  \frac{\vec{a}_n\cdot \hat{z}}{|n| L} \hat{z}^i.
\ee
which is now suppressed by $|n|L$. 

\subsubsection{Time-dependent coupling}
\label{sec:time-dependent-coupling}

It is interesting to see how the results change if we consider spherically symmetric radiation. 
For a structureless point particle, there is no natural way to do this. 
We can however model this situation (at least for a scalar field which has no gauge invariance) 
by considering a time dependent coupling constant $q$. 
This can be considered a toy model for a more realistic situation: 
for example, in a brane scenario a scattering event on the brane might lead to a spherically symmetric radiation of bulk modes. 
Alternatively, the coupling between the brane and $\phi$ 
might be determined by the vev of another field $\chi$ which could experience fluctuations.

In any case, consider the case where $q=q(t)$. The field is then
\be
\phi=\frac{1}{4\pi}\sum_n \frac{q(t)}{\sigma \cdot \dot{X}},
\ee
leading to a force
\be
F^i = \frac{1}{8\pi \gamma} \frac{\hat{z}^i}{|n| L \sqrt{1+\gamma^2 v_z^2} }\sum_n \frac{n}{|n|} \frac{\d}{\d t}\left(q^2\right)\Big|_{t_n}.
\ee
This now depends on the sign of $n$. 

It is interesting to consider the winding time -- 
the time it takes for an emitted photon to return to the particle. 
This is
\be
W_n \equiv |X^0(\tau_n) - X^0(0)| = 
\left| \frac{n \gamma L}{\frac{n}{|n|} \sqrt{1+\gamma^2 v_z^2} - \gamma v_z} \right| 
= |n| \gamma L \left( \sqrt{1+\gamma^2 v_z^2} + \frac{n}{|n|} \gamma v_z \right).
\ee
We see that an observer moving close to $c$ 
can experience many left-moving modes before a single right-moving mode. 
Furthermore, the sign of the effect is proportional to $\frac{\d q^2}{\d t}$, so the effect of, say, left moving modes can have either sign, either in the direction of motion or against the direction of motion of the particle.
We can therefore imagine an instability 
for a point particle moving at a speed close to $c$ relative to the topology frame that encounters many modes travelling in the same direction (that is, with the same $n/|n|$). Depending on the sign of the derivative of $q^2$, the impact could be in the same direction as the particle's velocity, which would either tend to increase the particle's speed toward $c$, or in the opposit direction, decreasing it to rest.

\subsection{Generalizations}
\label{sec:generalizations}
The results of the previous section can be extended to massless spin-1 and spin-2 bulk fields, 
and to arbitrary even space-time dimension with $D\geq 4$. 
The calculations are similar though more involved. 
Here we focus on the features that are new relative to the scalar case.

\subsubsection{Bulk Spin-1 Field}
The  action for a spin-1 field $A_\mu(x)$ is
\be
S  = \int \d^4 x \(-\frac{1}{4} F_{\mu\nu}^2\) + m \int \d \tau + q \int \d \tau \dot{X}^\mu A_\mu(X).
\ee
We proceed as above. The field $A_\mu$ can be computed in any gauge. 
Only the gauge invariant combination $F_{\mu\nu}=\partial_\mu A_\nu - \partial_\mu A_\nu$
enters into the point-particle equation of motion:
\be
	F_{\mu\nu} = 
		\frac{q}{4 \pi^2}
		\sum_n \frac{ 
			\Big[ (\sigma_n \cdot u) \ddot \epsilon(\bar{\tau}_n)_{[\mu} 
				- (\sigma_n \cdot \ddot \epsilon(\bar{\tau}_n)) u_{[\mu} \Big] 
			\partial_{\nu]} \bar{\tau}_n}
			{(\sigma_n \cdot u)^2}.
\ee
where $[a,b] = \frac{1}{2}(ab-ba)$.
We can then perturb the point-particle equation of motion
\be
	\ddot \ep^\mu 
		= \delta F^{\mu\nu}u_\nu + \bar{F}^{\mu\nu} \dot \ep_\nu 
		= \delta F^{\mu\nu} u_\nu.
\ee

Carrying through the above procedure we are lead to the force. 
In $D=4$ we get
\be
	F^i = 
		- \frac{q^2}{4\pi}
		\sum_n \frac{1}{|n|L \sqrt{1+\gamma^2 v_z^2}} 
			\left(
				\delta^{ij}- \frac{\hat{z}^i \left( \hat{z}^j + \gamma^2 v_z v^j  \right) }
								  {1+\gamma^2 v_z^2}   
			\right)a_n^j.
\ee

It is again interesting to consider some special cases. 
If $v||\hat{z}$, then $\sqrt{1+\gamma^2 v_z^2}=\gamma$. 
This leads to
\be
	F^i = 
		- \frac{q^2}{4\pi} 
		\sum_n 
		\frac{1}{|n| \gamma L} 
		\left(\delta^{ij} - \hat{z}^i\hat{z}^j \right) a_n^j.
\ee
Here we can see the dipole effect. 
In particular, when $\vec{v}||\vec{a}||\hat{z}$ there is zero force.
 
If instead $\vec{v}\bot \hat{z}$, then $\sqrt{1+\gamma^2 v_z^2}=1$. 
As a result,
\be
	F^i = 
	-\frac{q^2}{4\pi} 
	\sum_n \frac{1}{|n|L} 
	\left[\left(\delta^{ij} - \hat{z}^i \hat{z}^j \right)a_n^j \right].
\ee

\subsubsection{Bulk Spin-2 Field}
The action for a point particle coupled to General Relativity is
\be
S = \frac{\mpl^2}{2}\int \d^4 x \sqrt{-g} R + m \int \d \tau \sqrt{-g_{\mu\nu}\frac{\d X^\mu}{\d \tau}\frac{\d X^\nu}{\d \tau}}.
\ee
Expanding around flat space
\be
g_{\mu\nu} = \eta_{\mu\nu} + \frac{h_{\mu\nu}}{\mpl},
\ee
we can write the action to leading order in $h$ as
\be
S  = \int \d^4 x \(-\frac{1}{4} h_{\mu\nu} \mathcal{E}^{\mu\nu\rho\sigma} h_{\rho \sigma}\) + m \int \d \tau \left[1 + \frac{1}{2 \mpl} h_{\mu\nu}(X) \dot{X}^\mu \dot{X}^\nu \right],
\ee
where the Licherowicz operator is
\be
\E^{\mu\nu\rho\sigma} = \frac{1}{3!} \ep^{\mu\rho \alpha \lambda} \ep^{\nu \sigma \beta}_{\ \ \ \ \  \lambda}\partial_\alpha \partial_\beta.
\ee

Following the above logic we are lead in $D=4$ to
\ba
F^i &=& \frac{1}{4\pi}\left(\frac{m}{\mpl}\right)^2 \sum_n \frac{1}{|n| L \sqrt{1+\gamma^2 v_z^2}} \\
&& \times \left[a_n^i  - \frac{1}{4}\left( (\vec{a}_n\cdot \hat{z}) + \gamma^2 (\vec{a}_n\cdot \vec{v})(\vec{v} \cdot \hat{z}) \right) \left(\frac{1}{1+\gamma^2 v_z^2} \hat{z}^i + \frac{n/|n|}{\gamma \sqrt{1+\gamma^2 v_z^2}}v^i \right) \right].\nn 
\ea
This simplifies when $\vec{v} || \hat{z}$
\be
F^i(\vec{v} || \hat{z}) = -\frac{1}{4\pi} \left(\frac{m}{\mpl}\right)^2 \sum_n \frac{1}{|n| \gamma L} \left[ \delta^{ij}-\frac{1}{4}  \left(1+ \frac{n}{|n|} v \right) \hat{z}^i \hat{z}^j \right]a_n^j.
\ee
and when $\vec{v}\bot \hat{z}$
\be
F^i(\vec{v}\bot \hat{z}) = -\frac{1}{4\pi} \left(\frac{m}{\mpl}\right)^2 \sum_n \frac{1}{|n|  L} \left[ \delta^{ij}-\frac{1}{4} \hat{z}^i \hat{z}^j \right]a_n^j.
\ee

As a check, we should make sure that we can consistently ignore gravitational backreaction. This will be fine
\be
\Phi(\delta) \sim \frac{m}{M_{\rm Pl}^{2} \delta} \ll 1.
\ee
where $\delta$ is the regulated point particle's size. 
If $\delta \sim m^{-1} \ll L$ (so that the particle is not smaller than it's Compton wavelength), 
then this condition becomes $m \ll \mpl$.

In this case the size of an individual kick is
\be
\delta v_{\rm single\ kick,\ gravity}\ \sim \frac{M}{M_{\rm pl}}\frac{1}{(M_{\rm Pl} L)} \sim \Phi(L) < \Phi(\delta).
\ee
Thus the kicks are guaranteed to be small if we work in the probe limit. This analysis extends to arbitrary space-time dimensions.

\subsubsection{Other even space-time dimensions}
The above results generalize to any even space-time dimension. For the scalar and vector cases, the result only changes
by replacing
\be
\frac{q^2}{4\pi |n| L \sqrt{1+\gamma^2 v_z^2} } \rightarrow \frac{g^2}{2 (2 \pi)^{(D-2)/2} \left(|n| L \sqrt{1+\gamma^2 v_z^2}  \right)^{D-3} \Lambda^{D-4}}.
\ee
Here  $g$ is a dimensionless coupling constant
and $\Lambda$ is a scale associated with the interaction between the particle and field 
(which is non-renormalizable for $D>4$). 
The tensor structure is dimension-independent.

For the tensor case, since the propagator depends on $D$, the expresssion is slightly different. The net result is
\ba
F^i &=& \frac{1}{2(2\pi)^{(D-2)/2}}\frac{m^2}{\mpl^{{D-2}}} \sum_n \frac{1}{(|n| L \sqrt{1+\gamma^2 v_z^2})^{D-3}} \\
&& \times \left[a_n^i  - \frac{1}{2}\frac{D-3}{D-2}\left( (\vec{a}_n\cdot \hat{z}) + \gamma^2 (\vec{a}_n\cdot \vec{v})(\vec{v} \cdot \hat{z}) \right) \left(\frac{1}{1+\gamma^2 v_z^2} \hat{z}^i + \frac{n/|n|}{\gamma \sqrt{1+\gamma^2 v_z^2}}v^i \right) \right].\nn 
\ea

In all cases, an individual kick then has a size approximately given by 
\be
\delta v_{\rm single\ kick}\ \sim \frac{g^2}{(\Lambda L)^{D-3} M L}.
\ee
Within the regime of validity of our calculation, this will be small 
(assuming $g$ is small enough that we can trust perturbation theory).

Over the course of the trajectory, the particle can experience many such kicks,
however
\be
\sum^\infty_{n=-\infty} \frac{1}{|n|^{D-3}} \sim O(1) \ {\rm for\ }D>4.
\ee
Thus there is not a large accumulated effect for dimensions larger than four.

\section{Discussion}

In this work we have studied the self-force of a particle  in a compact space in a variety of contexts. 
The fact that compactifications break symmetries globally
can lead to position-dependent or velocity-dependent forces. 
In particular, winding modes of the field can create non-local and dissipative effects. 
For generic topologies that break translation invariance, such as the Klein bottle, 
the particle can generate a position dependent potential with which it interacts, 
creating privileged stable points in the space 
and preventing the particle from moving with constant velocity 
relative to the topology's natural rest frame.

Even for simple torus topologies, where translation invariance is preserved,
there can be effects due to the breaking of Lorentz invariance. 
A particle, once accelerated, will emit radiation 
that travels around the space and then interacts with the particle. 
We computed  the resulting force on the particle. 
This can be interpreted as 
a non-local correction to the particle's equation of motion due to the topology. 
When the particle is accelerated by an external force, 
the emitted radiation tends to act in the opposite direction of the original acceleration. 
In this sense the particle's trajectory is stable to the effects of winding modes. 
The effect can be large if there are a large number of fields in the bulk coupled to the particle.
A large effect can also build up over time if there are 4 space-time dimensions; 
such an effect may have occurred, for example, 
if in the early universe our four space-time dimensions were compact and small.

Intriguingly, for other sources of radiation, 
there can be an instability for particles moving close to $c$ in a closed space. 
We studied a toy example  by considering a particle with a time dependent coupling. 
In this case, the force exerted by right and left-moving modes is in opposite directions. 
However, a particle moving close to $c$ may experience many left-moving pulses (say) 
before a single right-moving pulse. 
Thus the particle can be pushed closer to $c$ or further from $c$, depending on the sign of $\dot{q^2}$.

Our results may be relevant in brane-world cosmology by changing the motion of the brane. 
There may also be particle physics implications. 
For example, the kicks may result in time dependence of the masses of the Kaluza-Klein modes. 
Finally, our results may also be relevant for some condensed matter systems 
where the system can be described with a closed topology.

The work in this paper could be further generalized. 
It may be interesting to extend the study to consider extended objects, not just point particles. 
The results can also be derived for more complex space forms, such as orientifolds or orbifolds, 
where we expect analogous effects to exist. 
It could also be interesting to study the effects in odd space-time dimensions 
by considering the tail effect. 
Finally, in this work we studied only radiation generated classically by a charged particle. 
It is also possible for an accelerated neutral particle to emit radiation quantum mechanically 
through the dynamical Casimir effect, so it would also be interesting to study the problem in that context.

\section*{Acknowledgements}
AAM is supported by an NSF GRFP. D.M. would like to thank CAPES for the fellowship Proc.8772-13-4, and also for the kind hospitality at Case Western Reserve University where this work was done.

\bibliographystyle{JHEPmodplain}
\bibliography{refs}

\end{document}